\documentclass[prl,twocolumn,showpacs,amsmath,amssymb,superscriptaddress]{revtex4}

\usepackage[dvips]{graphicx}
\usepackage{graphics}
\usepackage{dcolumn}
\usepackage{bm}
\usepackage[usenames]{color}

\begin{document}

\title{Experimental observation of the spin-Hall effect in a two dimensional spin-orbit 
coupled semiconductor system}

\author{J. Wunderlich}
\affiliation{Hitachi Cambridge Laboratory, Cambridge CB3 0HE, UK}

\author{B. K\"{a}stner}
\affiliation{Hitachi Cambridge Laboratory, Cambridge CB3 0HE, UK}
\affiliation{National Physical Laboratory, Teddington T11 0LW, UK}

\author{J. Sinova}
\affiliation{Department of Physics, Texas A\&M University, 
College Station, TX 77843-4242, USA}

\author{T. Jungwirth}
\affiliation{Institute of Physics ASCR, Cukrovarnick\'a 10, 162 53
Praha 6, Czech Republic} 
\affiliation{School of Physics and
Astronomy, University of Nottingham, Nottingham NG7 2RD, UK}

\date{\today}

\begin{abstract}
We report  the experimental observation  of the spin-Hall effect  in a
 two-dimensional  (2D) hole  system with  Rashba  spin-orbit coupling.
 The 2D  hole layer is a  part of a p-n  junction light-emitting diode
 with a  specially designed co-planar geometry which allows
 an angle-resolved polarization detection  at opposite edges of the 2D
 hole system. In  equilibrium the angular momenta of  the Rashba split
 heavy hole states lie in the plane of the 2D layer.  When an electric
 field  is applied  across the  hole channel  a non  zero out-of-plane
 component of the  angular momentum is detected whose  sign depends on
 the  sign  of  the  electric  field  and  is  opposite  for  the  two
 edges.  Microscopic  quantum  transport  calculations show only a
weak effect of disorder suggesting that the clean limit 
 spin-Hall conductance description (intrinsic spin-Hall effect)
 might apply to our system.
\end{abstract}


\maketitle  

The Hall effects  are among the most recognized  families of phenomena
in  basic  physics  and  applied microelectronics.  The  ordinary  and
quantum Hall effects, which,  e.g., proved the existence of positively
charged carriers (holes) in semiconductors and led to the discovery of
fractionally  charged quasiparticles \cite{Prange},  occur due  to the
Lorentz  force that  deflects {\em  like-charge} carriers  towards one
edge of  the sample creating a  voltage transverse to  the current. In
the anomalous Hall effect  \cite{HE}, the spin-orbit interaction plays
the role  of the force that  deflects {\em like-spin}  carriers to one
edge  and  opposite spins  to  the  other edge  of  the  sample. In  a
ferromagnetic material  this leads to  a net charge  imbalance between
the two sides which allows to detect magnetization in the conductor by
simple electrical means.  Here we report the experimental observation of
a new member  of the Hall family - the spin-Hall  effect (SHE).  As an
analogue  of the anomalous  Hall effect  but realized  in non-magnetic
systems  the  SHE  opens  new  avenues for  inducing  and  controlling
spin-currents  in semiconductors without  applying magnetic  fields or
introducing ferromagnetic elements.

Predictions  of  the SHE  were  reported, within different physical
contexts,  in  several seminal  studies
\cite{DP,Hirsch,Murakami,Sinova} and currently its microscopic origins
are  subject  of   an  intense  theoretical  debate  
\cite{Hu,Rashba,Schliemann,Shen,Inoue,Sinitsyn,Sheng,Nikolic,Schliemann2,Mischchenko,Raimondi,Nomura}.  
Experimentally, the SHE
has  been  elusive  because  in non-magnetic  systems  the  transverse
spin-currents do not  lead to net charge imbalance  across the sample,
precluding the simple electrical  measurement. To demonstrate the SHE,
we  have developed  a novel  p-n junction  light emitting  diode (LED)
microdevice, in a similar spirit to the one proposed in Ref. \onlinecite{Murakami}
but distinct in that it couples two dimensional hole and electron doped systems.  
Its co-planar  geometry and  the strong  spin-orbit (SO)
coupling  in the  embedded two-dimensional  hole gas  (2DHG),
whose  ultra-small
thickness deminishes current induced self-field effects,
 are well
suited for inducing and detecting  the SHE. 
 When an electric field is
applied across  the hole layer,  a non zero out-of-plane  component of
the spin is  optically detected whose sign depends on  the sign of the
field  and is  opposite  for  the two  edges,  consistent with  theory
predictions.

\begin{figure}[h]
\includegraphics[angle=0,width=3.5 in]{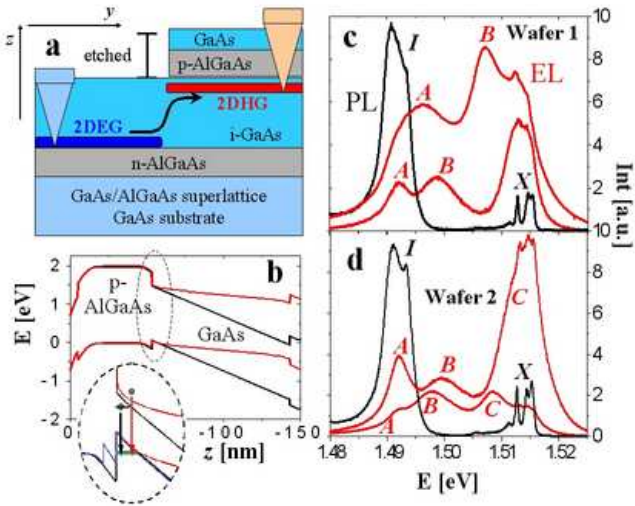} 
\caption{Basic   characteristics  of  the   LED.  (a)   The  schematic
cross-section  of the co-planar  p-n junction  LED device.  At forward
bias of  order of the GaAs  band-gap, electrons move from  the 2DEG to
the 2DHG  where they recombine.  The highest intensity of  the emitted
light is  in the p-region  near the junction step-edge.  (b) Numerical
simulations  of the  conduction and  valence band  profiles  along the
$\langle 001\rangle$  growth direction (z-axis) in  the un-etched part
of  the  wafer near  the  step edge.   Black  lines  correspond to  an
unbiased p-n junction and red lines  to a forward bias of 1.5 V across
the  junction.  In  the  detailed  image of  the  upper  (AlGa)As/GaAs
interface  (with  the  conduction  band  shifted down  in  energy  for
clarity) we  indicated possible sub-GaAs-gap  radiative recombinations
with the  2DHG involving  either 3D electrons  in the  conduction band
(red arrow) or impurity  states (black arrow).  Black line corresponds
to unbiased wafer 1 and blue line to wafer 2.  (c,d) Photoluminescence
spectra (black)  and electro-luminescence spectra (red)  at low (lower
curves)  and  high  (higher  curves)  bias measured  in  wafer  1  and
2 at 4.2 K. Impurity  ($I$),  excitonic  ($X$),  and  3D electron  to  2D  hole
transitions ($A$, $B$, $C$) are  identified.  We focus on peak B which
has small overlap with the $I$ and $X$ lines.}
\label{fig1}
\end{figure}

The   LEDs    were   fabricated   in    
(Al,Ga)As/GaAs   heterostructures grown by   molecular-beam epitaxy
and using modulation donor (Si) and acceptor (Be) doping in 
(Al,Ga)As barrier materials. The planar device features were prepared
by optical   and   electron-beam
lithography.  Schematic  of the  wafer  and  numerical simulations  of
conduction  and   valence  band  profiles  are   shown  in  
Fig.~\ref{fig1}(a,b)
\cite{Kastner}.  Two wafers were investigated  that differ in the 3 nm
undoped Al$_x$Ga$_{1-x}$As spacer at the  upper interface, with wafer 1 having
$x=0.5$  and  wafer  2  $x=0.3$.   The heterostructures  are  p-type;  the
band-bending  leads to the  formation of  an empty  triangular quantum
well in  the conduction  band at the  lower interface and  an occupied
triangular quantum well near the upper interface, forming a 2DHG.  The
co-planar p-n junction is created by removing acceptors from a part of
the wafer by etching, leading to population of the previously depleted
conduction band  well and  depletion of the  2DHG in that  region. The
diode  has a  rectifying $I-V$  characteristic  and the  onset of  the
current is accompanied by  electro-luminescence (EL) from the p-region
near  the junction  step edge  \cite{Kastner}.  The  current, $I_{LED}
\approx 100  \mu$A, is  dominated by electrons  moving from the  n- to
p-region; the opposite  hole current is negligibly small  due to lower
mobility of holes \cite{Kastner}.

The  interpretation of  EL peaks  shown  in Fig.~\ref{fig1}(c,d)  
starts from  a
comparison with measured photoluminescence (PL). Following previous PL
studies  \cite{Ossau,Silov} of  (Al,Ga)As/GaAs single-heterojunctions,
we assign  the sharp  high energy  PL peaks ($X$)  to well  known GaAs
exciton lines, and the lower energy peaks ($I$) to 2DEG to acceptor or
donor  to  2DHG  transitions.   The  other  PL  recombination  process
identified previously in  literature \cite{Ossau,Silov}, that involves
3D electron  to 2D hole transition  in the p-region  (or the analogous
process in the n-region), is  unlikely to contribute to our PL spectra
due to the large built-in electric field in the unbiased structure.

In  the EL  spectra,  the  excitonic peaks  are  clearly visible  and,
consistently, their  position is independent of wafer  and applied p-n
junction bias.   A non-zero EL  spectral weight at peaks  $I$ suggests
that recombination  processes with  impurity states may  contribute in
the low  energy part  of EL. The  peaks $I$  are expected to  shift to
lower energies under forward bias because of the weaker confinement at
the interfaces. Therefore, EL in the spectral range between the PL $I$
and $X$  peaks originates  from another mechanism,  i.e., from  the 3D
electron  to 2D hole  recombination. The  broad peak  $B$ is  the most
striking example of such a process.  Additional related peaks, $A$ and
$C$, can be  identified in wafer 1 at  high bias or in wafer  2 at low
bias,  respectively. These  EL  peaks shift  to  higher energies  with
increasing  $I_{LED}$, which  is  reminiscent of  the  behavior of  3D
electron to 2D  hole transition peaks as a  function of the excitation
power,  reported   in  the  PL  measurements   in  single  AlGaAs/GaAs
heterojunctions \cite{Ossau}.

\begin{figure}[h]
\includegraphics[angle=0,width=3.5 in]{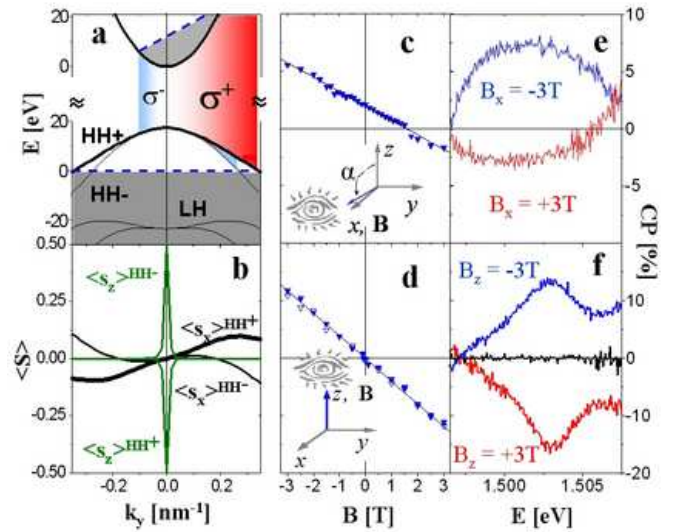}
\caption{Spin-polarization   in  the  absence   of  the   SHE  driving
current.  (a)  Theoretical  energy  spectrum  of  the  2DHG  along  ky
component of the wavevector (parallel to $I_{LED}$), of the heavy-hole
(HH)  and  light-hole  (LH)  subband,  spin-split due  to  the  Rashba
SO-coupling. Unbiased  wafer 1 was assumed in  the calculations. Small
shift of the  subband energies towards the top of  the valence band is
expected for wafer 2 and under bias. Schematic color plot demonstrates
circularly polarized  light emission  induced by transitions  from the
asymmetrically  occupied  conduction  band  to the  HH+  subband.  (b)
Theoretical spin-polarization  components along ky in the  HH+ and HH-
subbands.   (An  infenitisimal magnetic  field  along z-direction  was
added to  break the  degeneracy at ${\bf  k}=0$).  Except for  a small
region near $k_y=0$, spins  are oriented in-plane and perpendicular to
the  k-vector. (  c) Circular  polarization, defined  as  the relative
difference between intensities of  left and right circularly polarized
light, plotted as a function  of the in-plane magnetic field along the
x-axis, measured in  wafer 1. The observation angle  is $\alpha=85^o$ which is
the maximum  detection angle allowed  in our experimental  set-up. The
experiment measures,  to high accuracy,  the x-component of  the light
polarization   vector   which   is   proportional  to   $\langle   s_x
\rangle$. (d) Spectral  plot of the circular polarization  near peak B
for field $\pm$3 T. (e) Same  as (c) for magnetic and light detection axis
along the normal to the 2DHG plane. (f) Same as (d) for field 0 and $\pm$3
T.   At  zero  magnetic  field  the z-component  of  the  polarization
vanishes. All measurements are done at 4.2 K.}
\label{fig2}
\end{figure}

Fig.~\ref{fig2}(a)  presents our  microscopic  
calculations of  the 2DHG  energy
structure  \cite{Winkler}.  The  confining potential  of  the unbiased
wafer  1  was  assumed  in  these  numerical  simulations.   Only  one
heavy-hole  (HH) and  one light-hole  (LH)  bound state  forms in  the
quantum well.   The calculated spin  orientation as a function  of the
wavevector  component $k_y$ parallel  to the  $I_{LED}$ is  plotted in
Fig.~\ref{fig2}(b). 
Near ${\bf  k}=0$, the HH subband states  with total angular
momentum  $j_z=\pm 3/2$ are  split from  the LHs  with $j_z=\pm 1/2$  due to
different HH  and LH  effective masses, which  forces the HH  spins to
align with the z-axis. (Note  that in the 4-band representation of the
valence band states \cite{Winkler} the total angular momentum and spin
operators of  holes are related  as ${\bf j}$=3${\bf s}$.)  For larger
wavevectors, the mixing  with LH states together with  the Rashba type
of the  SO interaction in  the asymmetric quantum  well \cite{Winkler}
result in the  splitting of the HH subband and  in a vanishingly small
angular momentum component $\langle s_z\rangle$. The non-zero in-plane
component of  spin is oriented  perpendicular to the 2D  k-vector. For
the  measured hole density,  $p_{2D}=2\times 10^{12}$  cm$^{-2}$, only
the HH 2D subband is occupied by holes.

Assuming that  the optical recombination is dominated  by states close
the hole  subbands Fermi  wavevectors \cite{Silov} and  realizing that
the  3D  electron bands  occupation  under  forward  bias is  strongly
asymmetric along $k_y$, we can now assign peaks $A$  and $B$ to hole
transitions to the split HH subbands.  Peak $C$ is assigned to the LHs
which may  become populated  in biased  wafer 2 due  to a  weaker 2DHG
confinement, compared  to the one considered in  Fig.~\ref{fig2}. 
For simplicity
we  now focus  on peak  $B$  where the  role of  impurity and  exciton
recombinations can be safely neglected.

The  detection of  spin-polarization  phenomena is  done by  measuring
circular  polarization (CP)  of the  light. Due  to  optical selection
rules, a  finite CP along a  given direction of  the propagating light
indicates  a finite  spin-polarization in  this direction  of carriers
involved in  the recombination. In Fig.~\ref{fig2}(c,d)  
we plot the  CP of peak
$B$ for light detection axis close to the 2DHG plane and perpendicular
to $I_{LED}$. The  non-zero CP detected in the  absence of an external
magnetic  field,  consistent  with  the  results in  Fig.~\ref{fig2}(b),  
is  an
experimental  demonstration of a  spin-polarization effect  induced by
asymmetric  band occupation  \cite{Malshukov}.  Data  taken  at finite
magnetic  field  applied  along  the  x-direction  show  the  expected
\cite{Winkler} cooperative  or competing effects of the  field and the
Rashba SO  coupling, depending  on the sign  of the  field. Comparison
between these data  and data in Fig.~\ref{fig2}(e,f), 
showing  CP measurement at
magnetic  fields  and  light  detection axis  along  the  z-direction,
confirm the strongly anisotropic  Land\'e g-factor typical of the Rashba
SO coupled  2DHG \cite{Winkler}.  Fig.~\ref{fig2}(e,f)  
confirms that the {\em  CP is
purely in-plane  at zero  external magnetic field  which is  a crucial
piece  of evidence  in  the demonstration  of  the SHE},  which we  now
discuss.

\begin{figure}[h]
\includegraphics[angle=0,width=3.5 in]{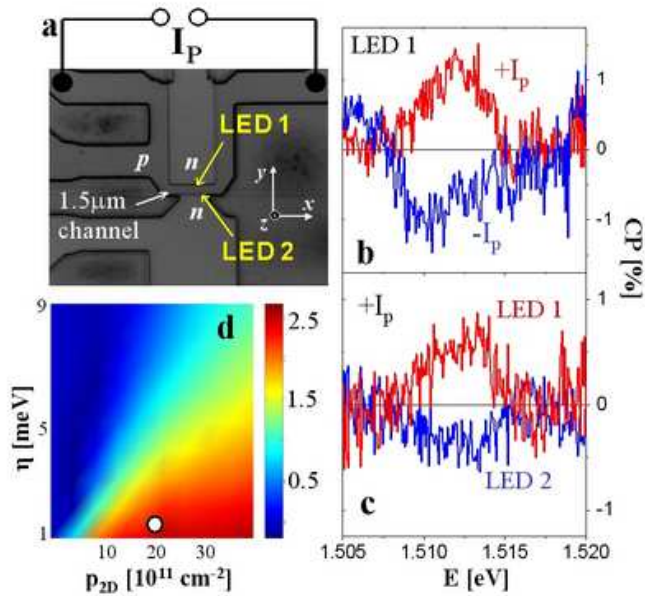}
\caption{SHE experiment. (a) Scanning electron microscopy image of the
SHE LED device.  The top (LED 1) or bottom (LED 2) n-contacts are used
to measure the EL at opposite  edges of the 2DHG p-channel parallel to
the SHE driving current  $I_p$. (b) Polarization along z-axis measured
with active LED  1 for two opposite Ip  current orientations. Spectral
region  of  peak  B  of  the   high  bias  EL  curve  of  wafer  1  is
shown.  Non-zero and  opposite out-of-plane  polarization for  the two
$I_p$ orientations demonstrates the SHE. (c) Polarization along z-axis
measured with fixed  $I_p$ current and for biased LED 1  or LED 2. The
data show opposite polarizations at opposite edges of the 2DHG channel
confirming  the SHE  origin of  the measured  signal.  (d) Theoretical
intrinsic  SHE conductivity  in  units of  $e/8\pi$ vs.  quasiparticle
life-time broadening and 2D hole density.  Parameters corresponding to
our 2DHG,  indicated by a white  dot, fall into the  strong i ntrinsic
SHE part of the theoretical diagram.}
\label{fig3}
\end{figure}

In the  SHE, non-zero  $\langle s_z \rangle$  occurs as a  response to
external  electric  field and  the  carries  with  opposite spins  are
deflected to opposite edges of  the sample parallel to the SHE driving
electrical current. A microdevice that  allows us to induce and detect
such  a response  is  shown  in Fig.~\ref{fig3}(a).  
A  p-channel current,  $I_p
\approx 100 \mu$ A, is applied  along x-direction and we measure CP of
the light  propagating along z-axis while  biasing one of  the LEDs at
either side of the p-channel.  The occurrence of the SHE upon applying
$I_p$ is demonstrated in Fig.~\ref{fig3}(b). 
When biasing, e.g., LED 1, peak $B$
polarization  of   the  light  emitted   from  the  region   near  the
corresponding  p-n  junction  step  edge  is non-zero  and  its  sign,
i.e.  the sign  of  $\langle s_z  \rangle$  of the  holes, flips  upon
reversing $I_p$.  The detected  polarization of 1\% represents a lower
bound of  the SHE  induced spin accumulation  at the edge  because the
emitted light intensity decreases  only gradually when moving from the
biased  junction  towards  the  opposite  side  of  the  2DHG  channel
\cite{Kastner}. To  highlight the consistency  of the signal  with the
unique  SHE phenomenology,  we  compare in  Fig.~\ref{fig3}(c)  
CPs obtained  in
experiment  where $I_p$  was  fixed and  either  LED 1  or  LED 2  was
activated.  The  opposite sign of these two  signals confirms opposite
$\langle s_z \rangle$ at the two edges, establishing the SHE origin.

To stimulate a detailed microscopic analysis of the observed
SHE, including the discussion of the role of disorder,
we present  in Fig.~\ref{fig3}(d) 
Kubo formula calculations  for wafer 1 of
the  spin Hall  conductivity, $\sigma_{SH}$,  which is  derived directly
from SO  coupled band structure and approaches  a disorder independent
value in high  mobility systems \cite{Murakami,Sinova,Schliemann}. For
the experimental density and  for a quasiparticle life-time broadening
of $1.2$  meV, estimated from  the measured mobility of  3400 cm$^2$Vs
and calculated HH  effective mass of 0.27 $m_e$, the  $\sigma_{SH}$ 
we obtain is
only weakly suppressed by disorder, suggesting that our sample
is in the regime referred to as the intrinsic SHE \cite{Murakami,Sinova}. 
Note that the robustness of this intrinsic SHE against weak disorder has
been challenged by several perturbation-theory analytical studies
\cite{Inoue,Mischchenko,Raimondi}. However,
the well defined clean limit of the SHE has recently been confirmed
by numerical quantum transport calculations based on exact eigenstates of
the disordered system, by numerical studies utilizing the Landauer-Buttiker
scattering formalism \cite{Sheng,Nikolic,Nomura,Hankiewicz}. In addition,
recent analytical calculations corresponding to our
2DHG confined in an asymmetric quantum well have shown that the vertex corrections
due to impurity scattering vanishes in this system, supporting the robustness
of the intrinsic SHE in the presence of weak disorder scattering in this system
\cite{SCZprivate}. 
Experimental exploration of  the phase diagram  in Fig.~\ref{fig3}(d) 
will help  to establish
the role of disorder in the observed effect.

We conclude by noting that,  beside the demonstrated ability to induce
and  detect  spin-polarization in  the  absence  of external  magnetic
fields, our  novel planar design  allows the fabrication of more complex
microdevices that  integrate on a single semiconductor  chip these two
functionalities  with  transmission of  the  spin information  between
different parts of a device.  Lithographically defined gate electrodes
may  provide additional  control  of  the SHE,  and  therefore of  the
overall device performance, by varying 2DHG density or SO coupling.

{\em Note added}: After our work, on which this Letter is based
on, was completed and presented at the 
Gordon Research Conference on Magnetic Nanostructures (Big Sky, Minessota, 
August 2004) and at the Oxford Kobe Seminar on Spintronics 
(Kobe, Japan, September 2004) an independent experimental observation
of the SHE has been reported \cite{Awschalom}. We point out that these
two experiments are in very different regimes, ours being in the strong
SO coupled regime in which the SO splitting is larger than the disorder
broadening, and the samples in Ref. \onlinecite{Awschalom} in the weak
SO coupled regime.

We thank Mohamed N. Khalid, Allan H. MacDonald, and Shoucheng Zhang for 
many useful discussions. The work was supported by Grant Agency of the Czech
Republic through Grant 202/02/0912.





\end{document}